\documentclass[prc,twocolumn]{revtex4-1}

\usepackage{enumerate}

\usepackage{booktabs}
\usepackage{color}
\usepackage{graphicx}

\usepackage{amsmath}
\usepackage{amssymb}
\usepackage{times}
\usepackage{multirow}

\def\bvec#1{\mbox{\boldmath $#1$}}

\newcommand{\del}[2]{\frac{\partial #1}{\partial #2}}

\newcommand{\bra}{\langle}
\newcommand{\ket}{\rangle}

\usepackage{ascmac}
\usepackage{fancybox}

\newcommand{\beq}{\begin{equation}}
\newcommand{\eeq}{\end{equation}}
\newcommand{\bea}{\begin{eqnarray}}
\newcommand{\eea}{\end{eqnarray}}

\def\fun#1#2{\lower3.6pt\vbox{\baselineskip0pt\lineskip.9pt
 \ialign{$\mathsurround=0pt#1\hfil##\hfil$\crcr#2\crcr\sim\crcr}}}

\begin{document}

\title{
  Analysis of the contribution of resonance poles of $^{16}$O based on the Mittag-Leffler theorem within the Jost-RPA framework
}

\author{K. Mizuyama$^{1,2}$}
\email{corresponding author: mizukazu147@gmail.com}
\author{T. Dieu Thuy$^{3,4}$}
\affiliation{
  \textsuperscript{1}
  Institute of Research and Development, Duy Tan University,
  Da Nang 550000, Vietnam
  \\
  \textsuperscript{2}
  Faculty of Natural Sciences,  Duy Tan University, Da Nang 550000, Vietnam
  \\
  \textsuperscript{3}
  Faculty of Physics, University of Education, Hue University,
  34 Le Loi Street, Hue City, Vietnam
  \\
  \textsuperscript{4} Center for Theoretical and Computational Physics,
  College of Education, Hue University, 34 Le Loi Street, Hue City, Vietnam
}

\date{\today}

\begin{abstract}
  This study investigates resonance states in $^{16}$O, specifically the $E$1 giant resonance and
  isoscalar/isovector quadrupole resonances, using the Jost-RPA method. By applying the Mittag-Leffler
  theorem, we decompose the RPA response function to analyze the individual contributions from poles
  corresponding to collective excitation modes. This decomposition clarifies the role of each pole in
  shaping the strength function, providing insights into the nature and classification of resonance
  states essential for understanding collective nuclear excitations. Our analysis identifies the dominant
  pole contributions to the observed resonance structures, offering a novel approach for exploring
  collective excitation phenomena in nuclei with complex many-body correlations.
\end{abstract}

\maketitle

\section{Introduction}
In finite quantum many-body systems, such as atoms and nuclei, resonance states are
characterized by peaks with widths in observables like cross sections or strength functions,
which are functions of energy.
Such resonance states have been studied theoretically for many years, mainly in S-matrix
theory~\cite{smat1,smat2}.
Resonance in quantum systems is understood as the phenomenon which occurs especially
in open systems, and is known as a state which decays in finite time~\cite{gamow1,gamow2}. 
This corresponds to the fact that the energy eigenvalues are complex, and the inverse of
the imaginary part of the complex eigenenergy is understood to give the lifetimes of the
resonant states. In S-matrix theory, complex energy eigenstates are given as poles of the
S-matrix. 
Also, the spatial behavior of the wavefunction of a decaying resonant state is very
different from that of a bound state, and it has the property of diverging outside
the atom or nucleus. This means that the poles of the S-matrix exist in the lower half
of the complex momentum plane, i.e., the imaginary part of the complex eigen momentum
is negative~\cite{Siegert}. 

The Jost function method~\cite{jost} is a method developed to compute the S-matrix and its poles
from a given Hamiltonian.
The Jost function is defined as a coefficient function of energy, linking the regular and irregular
solutions of a fundamental differential equation, such as the Schr\"{o}dinger equation, derived
from a Hamiltonian, 
and the poles of the
S-matrix can be found numerically as the zeros of the Jost function on the complex energy
plane. 
The Jost function method has been extended to multichannel systems to apply to realistic
problems and has been applied to atomic and nuclear
few-nucleon systems~\cite{Rakityansky1,Rakityansky2,Rakityansky}. 
In order to apply the Jost function method to the problem of the intermediate and heavier
mass nuclei, we have recently extended the Jost function method within the framework of the
Hartree-Fock-Bogoliubov (HFB) and
Random-Phase-Approximation (RPA) methods~\cite{jost-hfb,JostRPA,JostRPA2}. 

Since resonances in multi-channel systems are caused by complicated processes combining
many-body correlations, transitions to continuum, and interference, there are many unknowns
in the structure and mechanisms of resonances.
Resonances are typically observed as peaks in cross-sections or strength functions.
However, due to quantum interference with the continuum, the peak corresponding to a
resonance pole can disappear, as seen in Fano resonances~\cite{fano1,fano2,fano3,fano4,fano5,jost-fano,Litvinenko,jost-class}.
Conversely, peaks may appear even without a corresponding resonance pole. 
In nucleon scattering, the coupling between the incident nucleon and the collective excited
state of the target nucleus may produce sharp and narrow resonances. 
Collective excited states of nuclei observed in gamma excitation (e.g. giant resonance)
may also be resonances associated with the poles of the S-matrix.

In Ref.~\cite{JostRPA},
we extended the Jost function method within the RPA framework (Jost-RPA) and applied it to $E$1
giant resonances in $^{16}$O, finding poles corresponding to collective excitations (giant resonances).
However, just confirming the existence of a pole corresponding to the peak of the strength
function does not mean that one understands resonance. 
Therefore, in Ref.~\cite{JostRPA2},
we derived the S-matrix defined by the Jost function which satisfies unitarity
in the framework of RPA, and diagonalized the S-matrix to define the eigenphase shift. Then,
we further decomposed the RPA strength function into each eigenphase component. 
Applying these calculations to the quadrupole excitation of $^{16}$O, we found that the transition
density at the energy of the peak of the strength function, decomposed by the eigenphase shift
components, is characteristic of the collective excitation state. 
This indicates the possibility that the eigenphase-shift component decomposition can be
useful for classifying the excitation modes. 
However, the contribution of individual poles to the strength function is not clear yet,
and one cannot conclude whether the poles found in the complex energy plane by the Jost-RPA
method represent collective excitation modes without clarifying the individual contributions
of the poles. This is because the peak of the strength function may be formed by the superposition
of the contributions of several poles.

The purpose of this paper is to analyze the individual contributions of the poles corresponding
to the collective excitation modes in the $E$1 giant resonance and isoscalar/isovector
quadrupole resonances in $^{16}$O calculated in our previous papers~\cite{JostRPA,JostRPA2}. 
There are methods which allow the contributions of the poles to be extracted separately,
such as the complex scaling method~\cite{moiseyev,CSM1,CSM2} and the Gamow shell model~\cite{gamow-shell1,gamow-shell2}.

Following Berggren's approach~\cite{berggren}, we resolve pole contributions from the continuum by reconstructing
the system on the first Riemann sheet, adjusting the branch cut that connects it to other sheets.
However, since a method for applying these methods to the Jost-RPA method has not been
developed yet, we use the Mittag-Leffler theorem~\cite{Rakityansky,Jeffreys,Turner} as an
alternative to analyze the resonance pole contribution in this paper.

\section{Method}
In this section, we present a brief description of the Jost-RPA method,
the Riemann sheet defined in the basic equations of the Jost-RPA method
(RPA simultaneous second-order differential equations), and the use of the Mittag-Leffler theorem
used to resolve the resonance contributions in this paper.

\subsection{Jost-RPA method}
Second-order differential equations such as the time-independent Schr\"{o}dinger equation
have both regular and irregular solutions depending on the boundary conditions imposed
when solving the differential equation.
The Jost function is defined as the coefficient which connects the regular and irregular
solutions of a second-order differential equation as a function of complex energy or
momentum.
The Jost function can also be defined and calculated for such as the coupled channel
equation where the equation is given as a simultaneous differential equation.
The poles of the S-matrix which represent the resonance are given as the zeros of the
Jost function (or determinant of the Jost function matrix in the case of simultaneous
equations) on the complex energy plane.

In Refs.~\cite{JostRPA,JostRPA2}, the Jost function method was extended to the RPA method. 
The RPA equation is also a type of coupled channel equation and can be expressed as a
second-order simultaneous differential equation.
When there are $N$ configurations of RPA excitations, the RPA equations are expressed
as a simultaneous differential equation whose Hamiltonian operator is given by a
$2N\times 2N$ matrix as,
\begin{eqnarray}
  \left[
    -\frac{\hbar^2}{2m}
    \del{^2}{r^2}
    \bvec{1}_{2N}
    +
    \bvec{\mathcal{U}}
    +
    \delta
    \bvec{\mathcal{U}}
    \right]
  \vec{\bvec{\phi}}
  =
  \frac{\hbar^2}{2m}
  \bvec{\mathcal{K}}^2
  \vec{\bvec{\phi}}
  \label{RPAeq}.
\end{eqnarray}
In this RPA differential equation, $\bvec{1}_{2N}$, $\bvec{\mathcal{U}}$, $\delta\bvec{\mathcal{U}}$,
$\bvec{\mathcal{K}}$ and $\bvec{\Phi}$ are $2N\times 2N$ matrices and
can be represented by $N\times N$ block matrices as
\begin{eqnarray}
  &&
  \bvec{1}_{2N}
  =
  \begin{pmatrix}
    \bvec{1}_N & \bvec{0} \\
    \bvec{0} & \bvec{1}_N
  \end{pmatrix},
  \hspace{10pt}
  \bvec{\mathcal{K}}
  =
  \begin{pmatrix}
    \bvec{K}_1 & \bvec{0} \\
    \bvec{0} & \bvec{K}_2
  \end{pmatrix},
  \\
  &&
  \bvec{\mathcal{U}}
  =
  \begin{pmatrix}
    \bvec{U} & \bvec{0} \\
    \bvec{0} & \bvec{U}
  \end{pmatrix},
  \hspace{10pt}
  \delta
  \bvec{\mathcal{U}}
  =
  \begin{pmatrix}
    \delta\bvec{U} & \delta\bvec{U} \\
    \delta\bvec{U} & \delta\bvec{U}
  \end{pmatrix},
  \label{potterm}
\end{eqnarray}
where $\bvec{1}_N$ and $\bvec{U}$ are the $N$-dimensional unit matrix
and the diagonal matrix with the mean field potential, respectively.
$\bvec{K}_1$ and $\bvec{K}_2$ are complex momentum defined as $N$-dimensional diagonal matrices
with $k^{(1)}_{\alpha}=\sqrt{\frac{2m}{\hbar^2}(\epsilon_{\alpha}+E)}$ and
$k^{(2)}_{\alpha}=\sqrt{\frac{2m}{\hbar^2}(\epsilon_{\alpha}-E)}$
(with $\alpha\in 1,2,\cdots N$), respectively.
And $\epsilon_{\alpha}$ and $E$ are the hole state
and the complex excitation energy of the nucleus, respectively.
$\delta\bvec{U}$ is the $N\times N$ matrix of the residual
interaction defined by
\begin{eqnarray}
  \delta\bvec{U}(r)
  &=&
  \sum_{qq'}
  \bvec{\vec{\tilde{\varphi}}}_{q}(r)
  \frac{\kappa_{qq'}(r)}{r^2}
  \bvec{\vec{\tilde{\varphi}}}_{q'}^{\mathsf{T}}(r)
  \label{resV}
\end{eqnarray}
where 
$\bvec{\vec{\tilde{\varphi}}}$($\bvec{\vec{\tilde{\varphi}}}_n$ and
$\bvec{\vec{\tilde{\varphi}}}_p$) are hole state wavefunctions (for neutrons and protons)
defined as $N$-dimensional vectors.

By solving the RPA equation Eq.(\ref{RPAeq}) with boundary condition,
the solution $\vec{\bvec{\phi}}$ is obtained as a $2N$-dimensional vector as
\begin{eqnarray}
  \vec{\bvec{\phi}}(r)
  =
  \begin{pmatrix}
    \phi_1(r) \\
    \phi_2(r) \\
    \vdots\\
    \phi_\alpha(r) \\
    \vdots\\
    \phi_{2N}(r) \\
  \end{pmatrix}.
\end{eqnarray}
In general, for a $2N$-dimensional second-order simultaneous differential equation,
there exist $2N$ different ways of giving boundary conditions for regular and irregular
solutions, respectively, and they are linearly independent of each other.

The boundary condition for regular solution $\vec{\bvec{\phi}}^{(r\alpha)}(r)$ is given as
\begin{eqnarray}
  \lim_{r\to 0}
  \vec{\bvec{\phi}}^{(r\alpha)}(r)
  =
  \begin{pmatrix}
    0 \\
    0 \\
    \vdots\\
    F_{l_\alpha}(k_{\alpha}r) \\
    \vdots\\
    0
  \end{pmatrix}
\end{eqnarray}
so as to give the regularity of the solution near the origin at $r=0$, where
$k_{\alpha}$ is a diagonal component of $\bvec{\mathcal{K}}$ defined as $2N$-dimensional
diagonal matrix, and $F_{l}(kr)$ is a function defined by the spherical Bessel function as
$F_{l}(kr)\equiv rj_l(kr)$.

The boundary condition for irregular solutions $\vec{\bvec{\phi}}^{(\pm\alpha)}(r)$ is given as
\begin{eqnarray}
  \lim_{r\to \infty}
  \vec{\bvec{\phi}}^{(\pm\alpha)}(r)
  =
  \begin{pmatrix}
    0 \\
    0 \\
    \vdots\\
    O_{l_\alpha}^{(\pm)}(k_{\alpha}r) \\
    \vdots\\
    0
  \end{pmatrix}
\end{eqnarray}
so that the boundary condition for outgoing (or incoming) waves at infinity is satisfied,
where $O_{l}^{(\pm)}(kr)$ is a function defined by the spherical Hankel function as
$O_{l}^{(\pm)}(kr)\equiv rh_l^{(\pm)}(kr)$.

The solution matrix $\bvec{\Phi}$, given as a $2N\times 2N$ matrix, can be defined by
arranging the $2N$ solution vectors obtained with different boundary conditions as
\begin{eqnarray}
  \bvec{\Phi}
  \equiv
  \begin{pmatrix}
    \vec{\bvec{\phi}}^{(1)} & \vec{\bvec{\phi}}^{(2)} & \cdots & \vec{\bvec{\phi}}^{(2N)}
  \end{pmatrix}
  \label{Phimat}
\end{eqnarray}
and this satisfies
\begin{eqnarray}
  \left[
    -\frac{\hbar^2}{2m}
    \del{^2}{r^2}
    \bvec{1}_{2N}
    +
    \bvec{\mathcal{U}}
    +
    \delta
    \bvec{\mathcal{U}}
    \right]
  \bvec{\Phi}
  =
  \frac{\hbar^2}{2m}
  \bvec{\mathcal{K}}^2
  \bvec{\Phi}
  \label{RPAeq2}.
\end{eqnarray}
The Jost function is defined using the regular and irregular solution matrices
($\bvec{\Phi}^{(r)}$ and $\bvec{\Phi}^{(\pm)}$) as
\begin{eqnarray}
  &&
  \bvec{\mathcal{J}}^{(\pm)}
  \nonumber\\
  &&=
  \mp i
  \left[
    \bvec{\Phi}^{(r)\mathsf{T}}
    \left(
    \del{}{r}
    \bvec{\Phi}^{(\pm)}
    \right)
    -
    \left(
    \del{}{r}
    \bvec{\Phi}^{(r)\mathsf{T}}
    \right)
    \bvec{\Phi}^{(\pm)}
    \right]
  \bvec{\mathcal{K}}
  \nonumber\\
  \label{Jost}
\end{eqnarray}
as a coefficient function which connects the regular and irregular solution matrices as
\begin{eqnarray}
  \bvec{\Phi}^{(r)\mathsf{T}}
  &\equiv&
  \frac{1}{2}
  \left[
    \bvec{\mathcal{J}}^{(+)}
    \bvec{\Phi}^{(-)\mathsf{T}}
    +
    \bvec{\mathcal{J}}^{(-)}
    \bvec{\Phi}^{(+)\mathsf{T}}
    \right]
  \label{Jostdef}
\end{eqnarray}
The Green function defined as the $2N\times 2N$ matrix which satisfies
\begin{eqnarray}
  &&
  \left[
    \frac{\hbar^2}{2m}
    \bvec{\mathcal{K}}^2
    +\frac{\hbar^2}{2m}
    \del{^2}{r^2}
    \bvec{1}_{2N}
    -
    \bvec{\mathcal{U}}
    -
    \delta
    \bvec{\mathcal{U}}
    \right]
  \bvec{\mathcal{G}}^{(\pm)}(r,r')
  \nonumber\\
  &&
  =
  \delta(r-r')
  \bvec{1}_{2N}
  \label{RPAGreendef}
\end{eqnarray}
can be represented as
\begin{eqnarray}
  &&
  \bvec{\mathcal{G}}^{(\pm)}(r,r')
  \nonumber\\
  &&=
  \mp i
  \frac{2m}{\hbar^2}
  \left[
    \theta(r-r')
    \bvec{\Phi}^{(\pm)}(r)
    \bvec{\mathcal{K}}
    \bvec{\mathcal{J}}^{(\pm)-1}
    \bvec{\Phi}^{(r)\mathsf{T}}(r')
    \right.
    \nonumber\\
    &&
    \left.
    +
    \theta(r'-r)
    \bvec{\Phi}^{(r)}(r)
    \left(
    \bvec{\mathcal{J}}^{(\pm)-1}
    \right)^{\mathsf{T}}
    \bvec{\mathcal{K}}
    \bvec{\Phi}^{(\pm)\mathsf{T}}(r')
    \right].
  \label{green0}
\end{eqnarray}
The RPA response function is represented by using a Green function as
\begin{eqnarray}
  R_{qq'}(r,r';E)
  &=&
  \bvec{\vec{\xi}}_q^{\mathsf{T}}(r)
  \bvec{\mathcal{G}}^{(+)}(r,r';E)
  \bvec{\vec{\xi}}_{q'}(r')
  \label{RPAres}
\end{eqnarray}
where $\bvec{\vec{\xi}}_q$ is the $2N$-dimensional hole state vector $\bvec{\vec{\xi}}$
defined by using the $2N$-dimensional hole state vector $\bvec{\vec{\tilde{\varphi}}}_q$ as
\begin{eqnarray}
  \bvec{\vec{\xi}}_n
  \equiv
  \begin{pmatrix}
    \bvec{\vec{\tilde{\varphi}}}_{n} \\
    \bvec{\vec{\tilde{\varphi}}}_{n} 
  \end{pmatrix},
  \hspace{10pt}
  \bvec{\vec{\xi}}_p
  \equiv
  \begin{pmatrix}
    \bvec{\vec{\tilde{\varphi}}}_{p} \\
    \bvec{\vec{\tilde{\varphi}}}_{p} 
  \end{pmatrix}.
\end{eqnarray}
The RPA strength function is defined by using the RPA response function as
\begin{eqnarray}
  S_F(E)
  =
  -
  \mbox{ Im }
  R_{F}(E)
  \label{strdef}
\end{eqnarray}
with $R_{F}(E)$ defined by 
\begin{eqnarray}
  R_{F}(E)
  \equiv
  \frac{1}{\pi}
  \sum_{qq'}
  \int\int drdr'
  f_{q}^{\tau}(r)
  R_{qq'}(r,r';E)
  f_{q'}^{\tau}(r')
  \label{RFfunc}
\end{eqnarray}
where $f_{q}^{\tau}(r)$ is the external field which is defined by
\begin{eqnarray}
  f_{q}^{\tau=0}(r)
  &=&
  \delta_{qn}f(r)
  +
  \delta_{qp}f(r)
  \\
  f_{q}^{\tau=1}(r)
  &=&
  \delta_{qn}f(r)
  -
  \delta_{qp}f(r)
\end{eqnarray}
with $f(r)=r^L$ for Isoscalar ($\tau=0$) and Isovector ($\tau=1$), respectively.
For the case of the $E$1 dipole, the external field is given by
\begin{eqnarray}
  f_{q}^{E1}(r)
  &=&
  \delta_{qn}
  e\frac{Z}{A}
  f(r)
  -
  \delta_{qp}
  e\frac{N}{A}
  f(r).
\end{eqnarray}
\begin{figure}[htbp]
\includegraphics[width=\linewidth]{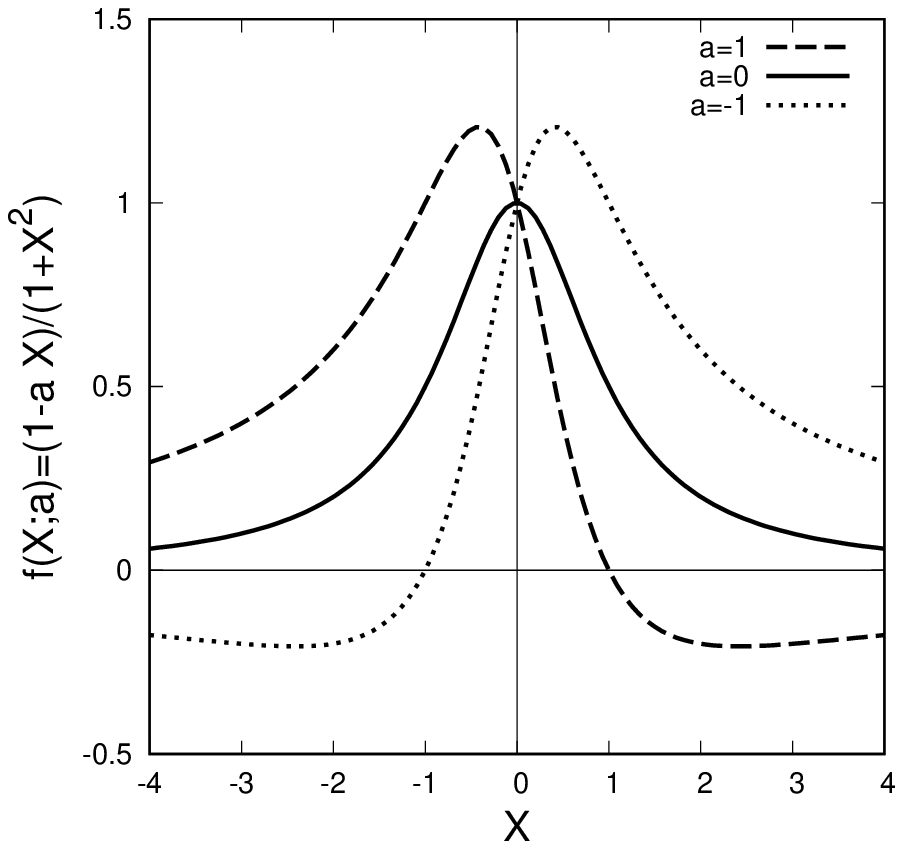}
\caption{
  The parameter ``$a$'' dependence of the function $f(X; a) = (1-aX)/(1+X^2)$ defined by Eq.(\ref{Fdef}).
  The graph shows the function for $a = 1$, $0$, and $-1$.
  }
\label{func}
\end{figure}
\subsection{Riemann sheet}
Due to the definition of the complex momentum, the complex momentum $k^{(1)}_\alpha(E)$
and $k^{(2)}_\alpha(E)$ ($\alpha\in 1,2,\cdots N$) are multivalued functions where two
values are determined for a single complex energy $E$. 
Conversely, when complex energy is considered as a function of complex momentum, 
the complex energy $E$ for which the imaginary part of all complex momentum is
positive is defined as a first Riemann sheet.

In other words, when all complex momentum $k^{(1)}_\alpha(E)$ and $k^{(2)}_\alpha(E)$
satisfy the symmetric property for a complex energy $E$ as,
\begin{eqnarray}
  k^{(1)}_{\alpha}(E^*)&=&-k^{(1)*}_{\alpha}(E),
  \\
  k^{(2)}_{\alpha}(E^*)&=&-k^{(2)*}_{\alpha}(E)
  \\
  &&
  \mbox{for all }\alpha(\in 1,2,\cdots,N)
  \nonumber
\end{eqnarray}
then we say that the complex energy $E$ belongs to the first Riemann sheet.
The poles $E_g$ obtained on the first Riemann sheet as solutions satisfying
$\det\bvec{\mathcal{J}}^{(+)}(E_g)=0$ using the Jost function are always
found in the upper half of the complex momentum ($k^{(1)}_\alpha$ and $k^{(2)}_\alpha$ for
all $\alpha$) plane and are solutions satisfying the boundary conditions of bound states. 

When the label $\alpha$ of the matrix element is defined so that energies of the hole
states $\epsilon_{\alpha}$(for $\alpha\in 1,2,\cdots,N$) are aligned as,
\begin{eqnarray}
  (0<)-\epsilon_{1}<-\epsilon_{2}<\cdots<-\epsilon_{N}
\end{eqnarray}
the $(c+1)$th Riemann sheet (for $c\in 1,2,\cdots,N$) in the positive energy region is
analytically connected to the first Riemann sheet on the real axis of complex energy $E$
within the range satisfying
$-\epsilon_{c} < \mbox{ Re } E < -\epsilon_{c+1}$
(when $c=N$, $-\epsilon_{N} < \mbox{ Re } E$), 
and forms a Riemann surface. This range on the real axis where the first Riemann sheet and
the $(c+1)$th Riemann sheet connect is called the branch line (or branch cut). 
On a Riemann surface formed by two connected Riemann sheets, the function of the
complex energy $E$ can be treated as a single-valued function. 

When one considers $E$ on the first Riemann sheet and $E^*$ on the $(c+1)$th Riemann sheet
across the branch line, the momentum $k^{(1)}_\alpha(E)$ and $k^{(2)}_\alpha(E)$ have the
following properties,
\begin{eqnarray}
  k^{(1)}_{\alpha}(E^*)
  &=&
  \left\{
  \begin{array}{lc}
    k^{(1)*}_{\alpha}(E) & \mbox{(for $1\leq\alpha\leq c$)} \\
    -k^{(1)*}_{\alpha}(E) & \mbox{(for $c+1\leq\alpha\leq N$)}
  \end{array}
  \right.
  \label{k1cprop}
\end{eqnarray}
and
\begin{eqnarray}
  k^{(2)}_{\alpha}(E^*)=-k^{(2)*}_{\alpha}(E)
  \hspace{3pt}(\mbox{for all }\alpha(\in 1,2,\cdots,N)).
  \label{k2prop}
\end{eqnarray}
In other words, the lower half of the complex $k^{(1)}_\alpha$-planes(for $1\leq\alpha\leq c$)
and the upper half of the complex $k^{(1)}_\alpha$-planes(for $c+1\leq\alpha\leq N$) and
$k^{(2)}_\alpha$-planes(for all $\alpha(\in 1,2,\cdots,N)$) correspond to the $(c+1)$th
Riemann sheet of the complex energy $E$.
The pole representing the resonance exists on the $(c+1)$th Riemann sheet, and the
position of the pole on the complex momentum plane corresponding to the Riemann sheet
gives the asymptotic behavior of the outgoing wave of the regular solution
$\bvec{\Phi}^{(r)}(r)$.
Since there is a relationship between $k^{(1)}_\alpha(E)$ and $k^{(2)}_\alpha(E)$ as
$k^{(1)}_\alpha(-E)=k^{(2)}_\alpha(E)$ by definition, there are the same number of
Riemann sheets which connect with the first Riemann sheet on the real axis in
the negative energy region. 
\begin{table}
  \caption{
    Information on the ground state of $^{16}$O and the branching structure of the
    Riemann sheet in the complex excitation energy plane. Hall state energies
    $\epsilon_\alpha$ (unit:MeV) and r.m.s. radii $\sqrt{\bra r^2\ket}$ (unit:fm)
    of neutrons and protons, and the energy range of the branching line with
    the first Riemann sheet in the positive energy region.
    }
  \label{table0}
  \begin{ruledtabular}
    \begin{tabular}{ccc}
       & Neutron & Proton \\
      \colrule
      & \multicolumn{2}{c}{hole states $\epsilon_\alpha$} \\
      $s_{1/2}$ & -36.17 & -31.16  \\
       $p_{3/2}$ & -21.31 & -16.84  \\
      $p_{1/2}$ & -16.38 & -11.95  \\
      \colrule
      & \multicolumn{2}{c}{r.m.s radius} \\
      $\sqrt{\bra r^2\ket}$ & 2.41 & 2.46 \\ 
    \end{tabular}
    \begin{tabular}{cccc}
      \multicolumn{4}{c}{Riemann sheets} \\
      \colrule
      &Sheet & Branch-cut [MeV]& \\
      \colrule
       &2nd & $11.95\leq E \leq 16.38$& \\
       &3rd & $16.38\leq E \leq 16.84$& \\
       &4th & $16.84\leq E \leq 21.31$& \\
       &5th & $21.31\leq E \leq 31.16$& \\
       &6th & $31.16\leq E \leq 36.17$& \\
       &7th & $36.17\leq E$& \\
    \end{tabular}
  \end{ruledtabular}
\end{table}
\subsection{Spectral expansion based on Mittag-Leffler theorem}
\begin{table}
  \caption{
    The poles $E_g$ of $E$1 excitation of $^{16}$O. The real part of the residue of $r_F^{(g)}$
    (Re $r_F^{(g)}$), the ratio of the imaginary to the real part
    ($\mbox{Im }r_F^{(g)}/\mbox{Re }r_F^{(g)}$) in Eq.(\ref{SFexpand2}),
    which gives the pole contribution of the strength function ($s_F^{(g)}(E)$).
    And the peak energy $E_{peak}$ of $s_F^{(g)}(E)$, calculated by Eq.(\ref{Epeak}).
  }
  \label{table1}
  \begin{ruledtabular}
    \begin{tabular}{cccrrc}
      Sheet & No. & $E_g$ & Re $r_F^{(g)}$ & $\frac{\mbox{Im }r_F^{(g)}}{\mbox{Re }r_F^{(g)}}$ & $E_{peak}$ \\
      & & [MeV] & [$\times 10^{-2} e^2fm^2$] & & [MeV] \\
      \colrule
      4th & (1) & $ 17.15-i0.02 $ & $ 0.01 $ & $ 0.50 $ & $17.14$ \\
          & (2) & $ 18.52-i0.72 $ & $ 0.17 $ & $-2.55 $ & $19.01$ \\
          & (3) & $ 19.28-i0.95 $ & $-0.17 $ & $ 2.96 $ & $18.60$ \\
          & {\bf (4)} & ${\bf 19.34-i0.53} $ & $ {\bf 9.36} $ & $ {\bf 0.19} $ & ${\bf 19.29}$ \\
          & (5) & $ 19.38-i0.97 $ & $ 0.32 $ & $ 5.95 $ & $18.56$ \\
          & {\bf (6)} & ${\bf 20.76-i0.34} $ & ${\bf 5.82} $ & ${\bf 0.26} $ & ${\bf 20.72}$ \\
      5th & {\bf (7)} & ${\bf 21.71-i0.63} $ & ${\bf 2.17}$ & ${\bf -5.66} $ & ${\bf 22.24}$ \\
    \end{tabular}
  \end{ruledtabular}
\end{table}
\begin{figure}[htbp]
  \includegraphics[width=\linewidth]{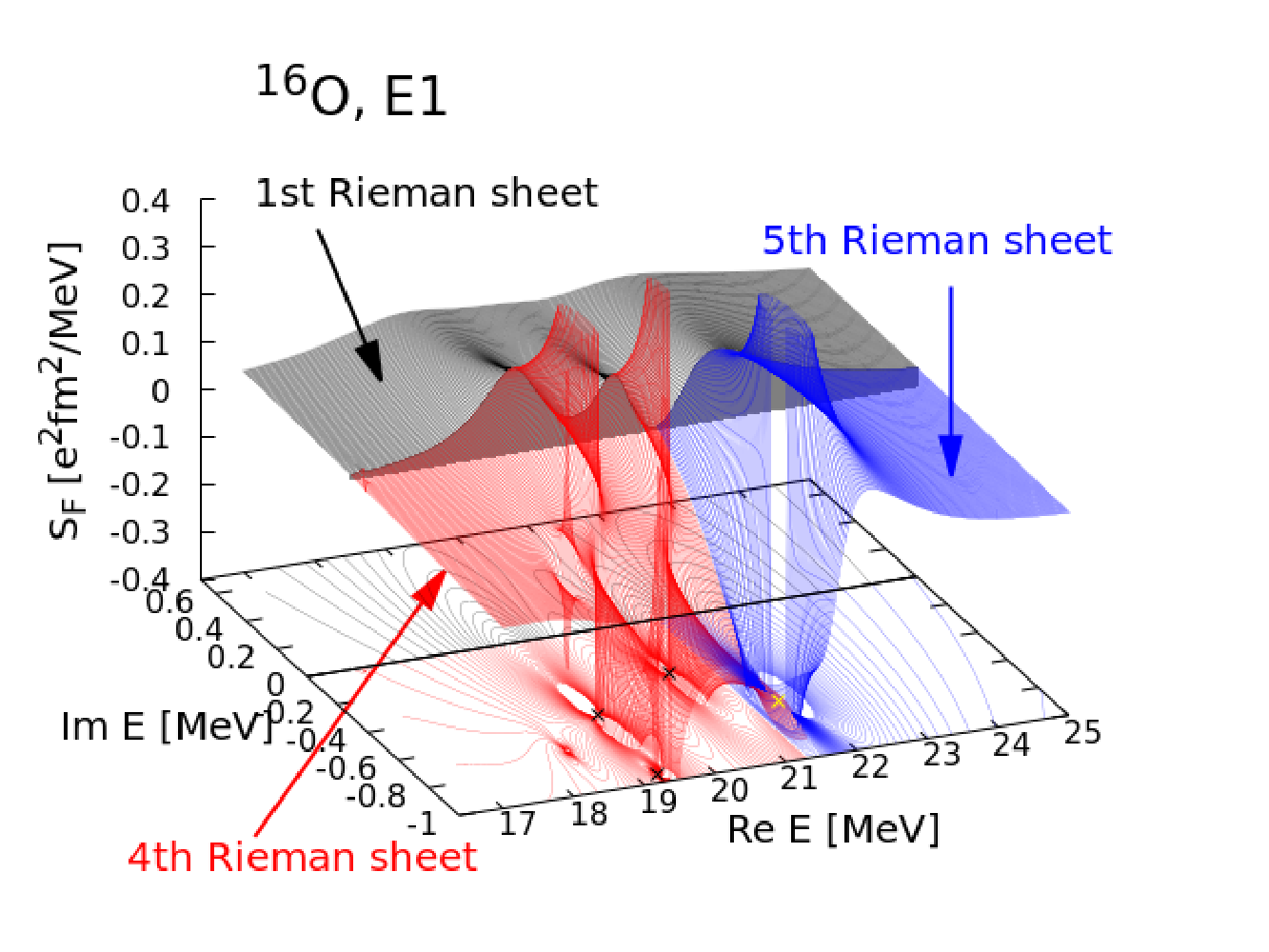}
  \includegraphics[width=\linewidth]{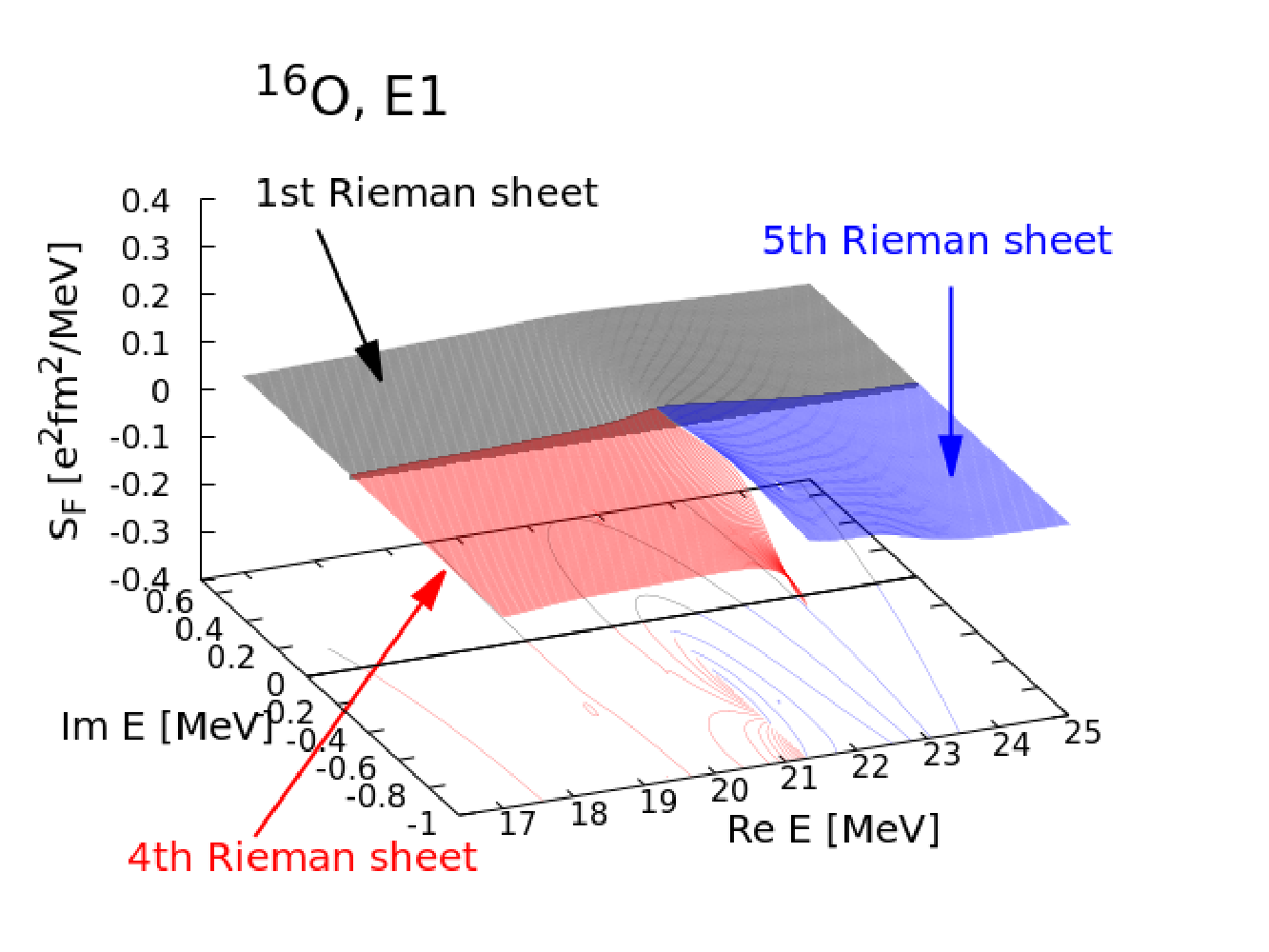}
  \caption{(Color online)
    RPA strength function for $E$1 excitation of $^{16}$O shown on
    the Riemann surface of complex excitation energy $E$.
    The lower panel shows the regular component of the strength function
    (the first term on the right-hand side of Eq.(\ref{SFexpand})), where
    the resonance pole contributions have been removed from the upper panel.
  }
  \label{srtGDR3D}
\end{figure}
In complex analysis, a function which is defined in a certain domain of the Riemann surface,
has no singularities other than poles, and is regular except at singular points,
is called a meromorphic function.
According to the Mittag-Leffler theorem~\cite{Rakityansky,Jeffreys,Turner},
the meromorphic function $f(z)$ which has
poles can be decomposed into a sum of the parts giving the pole contribution and other
regular (background) part $f^{reg}(z)$ as
\begin{eqnarray}
  f(z)
  =
  f^{reg}(z)
  +
  \sum_{g}
  \frac{1}{z-z_g}\mbox{Res}[f,z_g].
  \label{MLtheorem}
\end{eqnarray}

The RPA response function $R_F(E)$ (Eq.(\ref{RFfunc})) is a meromorphic function
which is defined on the Riemann surface of the complex energy $E$ and has poles.
If we define the residue of $R_F$ as
\begin{eqnarray}
  r_F^{(g)}
  \equiv
  \mbox{Res}[R_F,E_g]
  =
  \oint_{C_g}\frac{dE'}{2\pi i}R_F(E')
  \label{rfdef}
\end{eqnarray}
and use the Mittag-Leffler theorem (Eq.(\ref{MLtheorem})),
$R_F(E)$ can be expressed as
\begin{eqnarray}
  R_F(E)
  &=&
  R_F^{reg}(E)
  +
  \sum_{g}
  \frac{r_F^{(g)}}{E-E_g}
  \label{RFexpand}
\end{eqnarray}
where $R_F^{reg}(E)$ is the regular term obtained by removing the pole contribution from $R_F$. 

By inserting Eq.(\ref{RFexpand}) into the definition of the RPA strength function $S_F(E)$
Eq.(\ref{strdef}), $S_F(E)$ can also be decomposed into the regular term and the
pole contribution as
\begin{eqnarray}
  S_F(E)
  &=&
  S_F^{reg}(E)
  +
  \sum_{g}
  s_F^{(g)}(E)
  \label{SFexpand}
\end{eqnarray}
with
\begin{eqnarray}
  &&
  S_F^{reg}(E)
  =
  -
  \mbox{ Im }
  R_{F}^{reg}(E)
  \label{SFexpand1}
  \\
  &&
  s_F^{(g)}(E)
  =
  -
  \mbox{ Im }
  \left(
  \frac{r_F^{(g)}}{E-E_g}
  \right)
  \nonumber\\
  &&=
  \left(
  \frac{\mbox{ Re }r_F^{(g)}}{-\mbox{Im }E_g}
  \right)
  f\left(\frac{E-\mbox{Re }E_g}{-\mbox{Im }E_g};\frac{\mbox{Im }r_F^{(g)}}{\mbox{Re }r_F^{(g)}}\right)
  \label{SFexpand2}
\end{eqnarray}
where $f(X,a)$ is defined by
\begin{eqnarray}
  f(X;a)
  \equiv
  \frac{1-a X}{1+X^2}.
  \label{Fdef}
\end{eqnarray}
Fig.\ref{func} shows the dependence of the function $f(X,a)$ on the parameter $a$.
The function $f(X,a)$ has a so-called Breit-Wigner type symmetric shape with a peak
at X=0 when $a=0$. However, when the parameter $a$ has a finite value, the shape
of the function becomes asymmetric.

The peak of the function $f(X,a)$ is given by
\begin{eqnarray}
  f(X_{peak},a)=\frac{\sqrt{1+a^2}}{2\left\{1-\frac{\sqrt{1+a^2}-1}{a^2}\right\}}
  \hspace{5pt}
  (\geq 1)
  \label{fpeak}
\end{eqnarray}
with
\begin{eqnarray}
  X_{peak}=\frac{1-\sqrt{1+a^2}}{a}
  \label{Xpeak}
\end{eqnarray}
Note that, when taking the limit of $a\to 0$,
Eq.(\ref{fpeak}) becomes 1 because $\lim_{a\to 0}\frac{\sqrt{1+a^2}-1}{a^2}=1/2$,
and Eq.(\ref{Xpeak}) becomes $0$ at the limit of $a\to 0$, and
the integral value of this function is also given by
\begin{eqnarray}
  \int_{-\infty}^{\infty} dX f(X,a)
  =\pi
  \label{intf}.
\end{eqnarray}
Due to these properties about the function $f(X,a)$, the contribution of the pole
of the strength function, $s_F^{(g)}(E)$, has a peak at
\begin{eqnarray}
  E=\mbox{Re }E_g
  -
  \left[
    \frac{1-\sqrt{1+\left(\frac{\mbox{Im }r_F^{(g)}}{\mbox{Re }r_F^{(g)}}\right)^2}}
         {\left(\frac{\mbox{Im }r_F^{(g)}}{\mbox{Re }r_F^{(g)}}\right)}
         \right]
  \mbox{Im }E_g
  \label{Epeak}
\end{eqnarray}
and its amplitude is proportional to $r_F^{(g)}$ and inversely proportional to $-\mbox{Im }E_g$.
And, the deviation of $\mbox{Im }r_F^{(g)}/\mbox{Re }r_F^{(g)}$ from $0$ makes the shape of
$s_F^{(g)}(E)$ deviate from the Breit-Wigner type shape and become asymmetric.

From the integral property of the function $f(X,a)$ as shown in Eq.(\ref{intf}),
the energy integral of $s_F^{(g)}(E)$ is given by
\begin{eqnarray}
  \int_{-\infty}^{\infty} dE
  s_F^{(g)}(E)
  =
  \pi
  \mbox{Re }r_F^{(g)}.
  \label{intsF}
\end{eqnarray}
Therefore, it can be seen that $\mbox{Re }r_F^{(g)}.$ represents the substantial contribution
to the strength function, regardless of $\mbox{Im }r_F^{(g)}$ or $\mbox{Im } E_g$.

Defining the density fluctuation $\delta\rho_{F,q}(r;E)$ as,
\begin{eqnarray}
  \delta\rho_{F,q}(r;E)
  \equiv
  \frac{1}{\pi}
  \sum_{q'}
  \int dr'
  R_{qq'}(r,r';E)
  f_{q'}^{\tau}(r')
  \label{trdf}
\end{eqnarray}
since this is also a meromorphic function on the Riemann surface of complex energy $E$,
$\delta\rho_{F,q}(r;E)$ can also be expanded as
\begin{eqnarray}
  \delta\rho_{F,q}(r;E)
  &=&
  \delta\rho_{F,q}^{reg}(r;E)
  +
  \sum_g
  \delta\rho_{F,q}^{(g)}(r;E)
  \label{trdf-exp}
\end{eqnarray}
where
\begin{eqnarray}
  \delta\rho_{F,q}^{(g)}(r;E)
  \equiv
  \frac{1}{E-E_g}
  \mbox{Res}\left[\delta\rho_{F,q}(r),E_g\right].
  \label{trdpole}
\end{eqnarray}
These density fluctuations are related with the strength function as
\begin{eqnarray}
  &&
  S_F(E)
  =
  -
  \sum_q
  \int dr
  f_q^\tau(r)
  \mbox{ Im }
  \delta\rho_{F,q}(r;E)
  \label{trdexpand0},
  \\
  &&
  S_F^{reg}(E)
  =
  -
  \sum_q
  \int dr
  f_q^\tau(r)
  \mbox{ Im }
  \delta\rho_{F,q}^{reg}(r;E)
  \label{trdexpand1},
  \\
  &&
  s_F^{(g)}(E)
  =
  -
  \sum_q
  \int dr
  f_q^\tau(r)
  \mbox{ Im }
  \delta\rho_{F,q}^{(g)}(r;E)
  \label{trdexpand2}.
\end{eqnarray}
Hereafter, when we refer to ``density fluctuations'' in this paper,
we refer to the integrand part of the strength function described above.
\begin{figure}[htbp]
  \includegraphics[width=\linewidth]{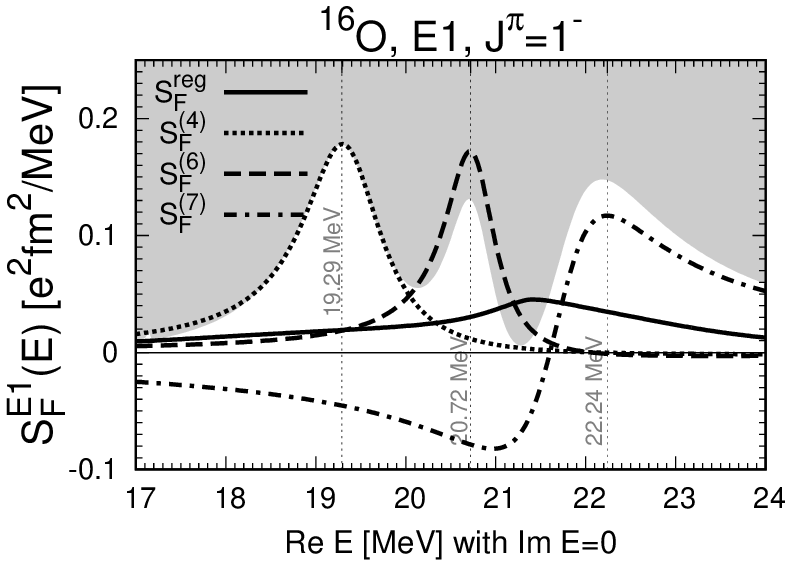}
  \caption{
    Spectral expansion of the RPA strength function for $E$1 excitation of $^{16}$O using Eq.(\ref{SFexpand}).
    The $S_F^{reg}$ shown by the solid curve is the regular component (the first term on the right side of Eq.(\ref{SFexpand}))
    obtained by removing the resonance pole contribution shown in Table \ref{table1} from the total strength function.
    $S_F^{(4)}$, $S_F^{(6)}$, and $S_F^{(7)}$ (dotted, dashed, and single-dotted curves) show the contributions of poles (4), (6), and (7) to
    the strength functions shown in Table \ref{table1}, respectively.  
    The dotted perpendicular lines represent the $E_{peak}$ positions ($19.29$, $20.72$, and $22.24$ MeV)
    for (4), (6), and (7), respectively.
  }
  \label{strpoleE1}
\end{figure}

\begin{figure}[htbp]
  \includegraphics[width=\linewidth]{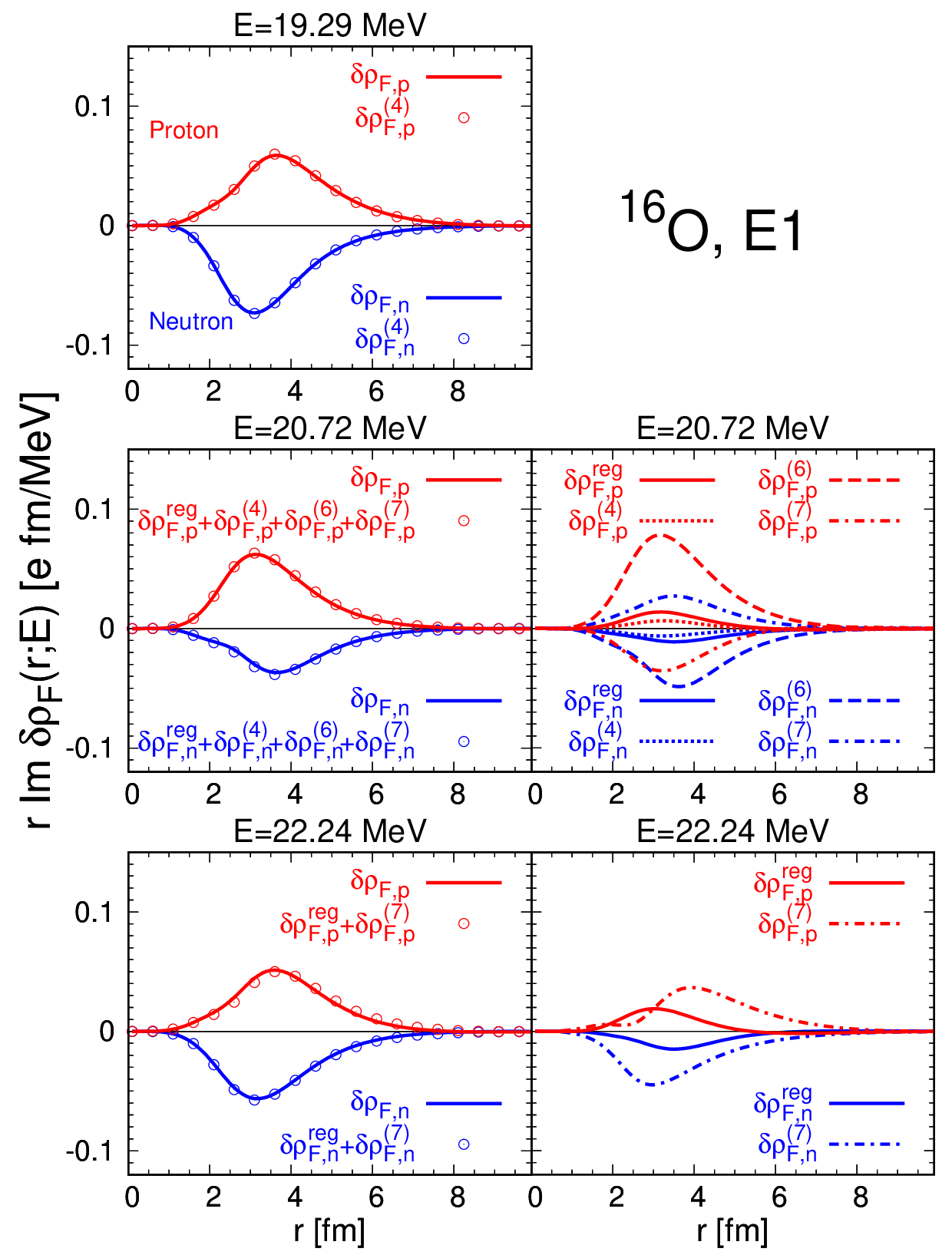}
  \caption{(Color online)
    Density fluctuations for $E$1 excitations of $^{16}$O defined by Eq.(\ref{trdf}) at $E=19.29$, $20.72$, and $22.24$ MeV.
    The neutron and proton components of the density fluctuations are shown as blue and red curves, respectively.
    In the left panels, the main terms forming the density fluctuation $\delta\rho_{F,q}(r;E)$ are compared with the
    total $\delta\rho_{F,q}(r;E)$, and in the right panels, each component decomposed according to Eq.(\ref{trdf-exp}) is shown.
  }
  \label{trdE1}
\end{figure}
\begin{table*}
  \caption{
    The poles $E_g$ of isoscalar/isovector quadrupole excitation of $^{16}$O. The real part of the residue of $r_F^{(g)}$
    (Re $r_F^{(g)}$), the ratio of the imaginary to the real part
    ($\mbox{Im }r_F^{(g)}/\mbox{Re }r_F^{(g)}$) in Eq.(\ref{SFexpand2}),
    which gives the pole contribution of the strength function ($s_F^{(g)}(E)$).
    And the peak energy $E_{peak}$ of $s_F^{(g)}(E)$, calculated by Eq.(\ref{Epeak}).
  }
  \label{table2}
  \begin{ruledtabular}
    \begin{tabular}{cccrrcrrc}
      & & & \multicolumn{3}{c}{Isoscalar $2^+$} & \multicolumn{3}{c}{Isovector $2^+$} \\
      \cline{4-6} \cline{7-9}
      Sheet & No. & $E_g$ & Re $r_F^{(g)}$ & $\frac{\mbox{Im }r_F^{(g)}}{\mbox{Re }r_F^{(g)}}$ & $E_{peak}$& Re $r_F^{(g)}$ & $\frac{\mbox{Im }r_F^{(g)}}{\mbox{Re }r_F^{(g)}}$ & $E_{peak}$ \\
      &&[MeV] & [$\times 10^{-2} fm^4$] & & [MeV]  & [$\times 10^{-2} fm^4$] & & [MeV] \\
      \colrule
      3th & (1) & ${\bf 16.76-i0.11} $ & ${\bf 2.64\times 10^3}$ & ${\bf -0.05} $ & ${\bf 16.76}$ & $ -0.89 $ & $ -2.72 $ & $16.84$ \\
      4th & (2) & $ 18.74-i0.68 $ & $    0.09 $ & $-0.56 $ & $18.91$ & $  0.25 $ & $ -7.36 $ & $19.33$ \\
          & (3) & $ 18.74-i0.71 $ & $    2.17 $ & $-1.19 $ & $19.07$ & $  0.15 $ & $-13.29 $ & $19.40$ \\
      5th & {\bf (4)} & ${\bf 27.02-i1.50} $ & ${\bf -4.77} $ & ${\bf-1.93} $ & ${\bf 27.93}$ & ${\bf 2.58\times 10^2}$ & ${\bf -0.67} $ & ${\bf 27.47}$ \\
          & {\bf (5)} & ${\bf 28.48-i0.57} $ & ${\bf -2.74} $ & ${\bf 0.95} $ & ${\bf 28.25}$ & ${\bf 11.26} $ & ${\bf -5.12} $ & ${\bf 28.95}$ \\
          & (6) & $ 30.05-i0.66 $ & $   -0.72 $ & $-0.23 $ & $30.12$ & ${\bf 9.45} $ & ${\bf -16.58} $ & ${\bf 30.67}$ \\
      6th & (7) & $ 33.63-i0.91 $ & $   -0.07 $ & $ 7.13 $ & $32.83$ & $ -3.62 $ & $ -0.25 $ & $33.74$ \\
          & {\bf (8)} & ${\bf 34.75-i0.59} $ & ${\bf 6.97} $ & ${\bf 0.49} $ & ${\bf 34.61}$ & ${\bf 17.34} $ & ${\bf -2.44} $ & ${\bf 35.14}$ \\
      7th & {\bf (9)} & ${\bf 36.61-i0.42} $ & ${\bf 2.71} $ & ${\bf 0.49} $ & ${\bf 36.51}$ & ${\bf 13.85} $ & ${\bf -5.83} $ & ${\bf 36.96}$ \\
    \end{tabular}
  \end{ruledtabular}
\end{table*}

\begin{figure}[htbp]
\includegraphics[width=\linewidth]{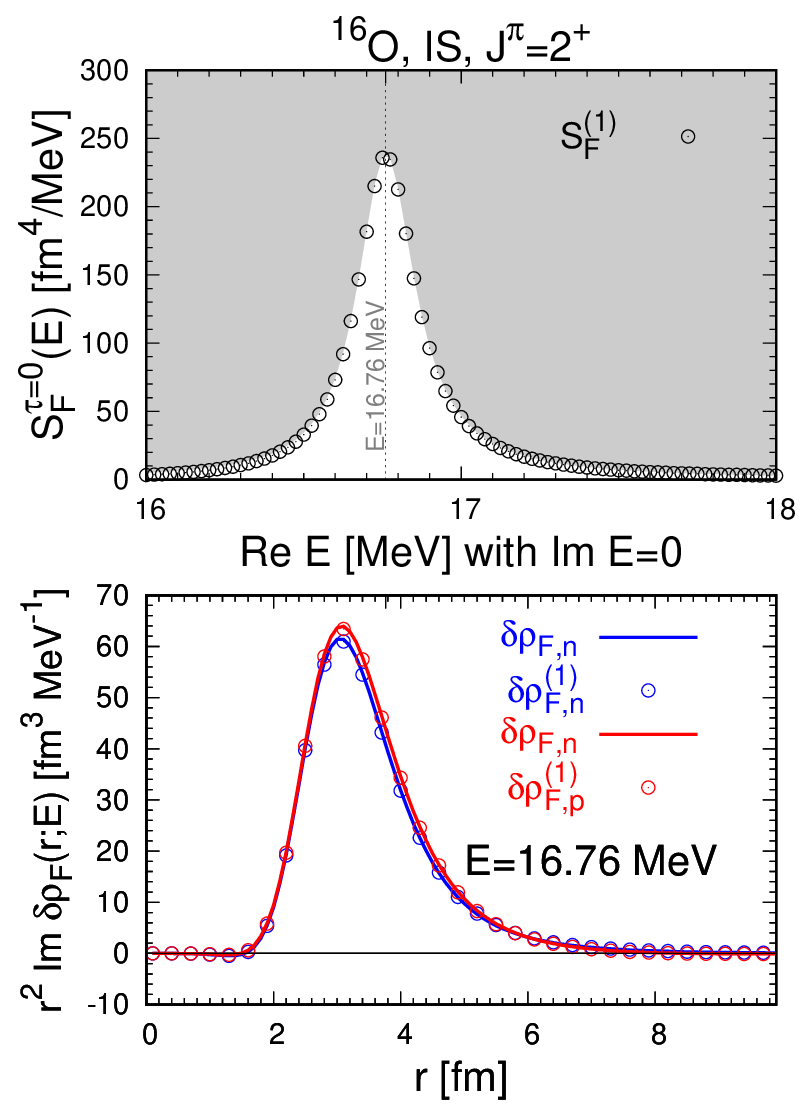}
\caption{(Color online)
  The strength function near $E = 16.36$ MeV (top panel) and
  density fluctuation at near $E = 16.36$ MeV(bottom panel) of
  the isoscalar quadrupole excitation of $^{16}$O.
}
\label{strpoleL2-1}
\end{figure}

\section{Results and discussions}
In this paper, as in Refs. \cite{JostRPA,JostRPA2}, we chose the Woods-Saxon potential
and residual interaction with the same parameters for Eq.(\ref{potterm}),
and $^{16}$O as the target nucleus.
The analysis is performed in the $E$1 and quadrupole excitation modes of $^{16}$O,
with RPA configuration numbers $N = 14$ ($N_n = 7$ and $N_p = 7$) and $N = 16$
($N_n = 8$ and $N_p = 8$) for $E$1 and quadrupole excitation, respectively
(see Table II in Ref.\cite{JostRPA} and Table II in Ref.\cite{JostRPA2}, respectively).

In Table \ref{table0}, the ground state information (hole state energies $\epsilon_\alpha$
and mean square radius $\sqrt{\bra r^2\ket}$) of $^{16}$O, and the branching lines of
the Riemann sheets which connect to the first Riemann sheet on the real axis of energy
in the positive energy region defined by the hole state energies $\epsilon_\alpha$ and
momentum $k_{\alpha}^{(1)}$, are shown.

\subsection{$E$1 dipole excitations}
In this section, using Eq.(\ref{SFexpand})-(\ref{SFexpand2}), we decompose the pole contributions and analyze the pole
contributions to the total strength function numerically for $E$1 excitation and isoscalar,
isovector quadrupole excitation in $^{16}$O.
In Table \ref{table1}, we show the poles $E_g$ found on the Riemann sheet, the real part of the
residue Re $r_F^{(g)}$, the ratio of the imaginary to the real part Im $r_F^{(g)}$/Re $r_F^{(g)}$,
and the peak energy $E_{peak}$ calculated using Eq.(\ref{Epeak}). 

The regular component $S_F^{reg}(E)$ of the strength function in the first term of the
right-hand side of Eq.(\ref{RFexpand}) can be obtained by removing the pole contribution
(second term of the right-hand side) from the total strength function $S_F(E)$ on
the left-hand side. 
In order to demonstrate that the pole contribution is removed from the strength
function $S_F(E)$ by Eq.(\ref{SFexpand}), $S_F(E)$ is shown in the upper panel 
and $S_F^{reg}(E)$ is shown in the lower panel of Fig.\ref{srtGDR3D}, 
where $S_F^{reg}(E)$ is calculated by subtracting from $S_F(E)$ the contributions of
all resonant poles on the 4th and 5th Riemann sheets shown in Table \ref{table1}.
Contour curves of the strength function are shown at the bottom of the figure to
clearly illustrate the analytical continuity between different Riemann sheets. 

In the upper panel, the strength function calculated with $E$ belonging to the
first Riemann sheet is plotted in the region of Im $E\geq 0$. 
In the Im $E\leq 0$ region, the strength functions were calculated with $E$ belonging
to the 4th and 5th Riemann sheets and plotted in the regions $16.84\leq \mbox{Re } E \leq 21.31$
MeV and $21.31\leq \mbox{Re } E \leq 31.16$ MeV, respectively. 
The strength functions computed on the 4th and 5th Riemann sheets in the region
Im $E\leq 0$ are not connected to each other.
However, both the strength functions on the 4th and 5th Riemann sheets are smoothly
and analytically connected to the 1st Riemann sheet on the real axis. 
By plotting the strength function in this way, we can see how the poles propagate and
affect the strength function on the real axis in the upper panel. 
We can also see that some poles have little effect on the shape of the strength
function on the real axis.
Not all of the poles in the Im $E\leq0$ region affect the shape of the strength function
on the real axis, and some poles have little or no effect.

The lower panel shows the strength function obtained by removing the pole contributions
shown in Table \ref{table1} from the strength function shown in the upper panel.
The divergence behavior of the strength function due to the existence of poles seen in
the region Im $E \leq 0$ has disappeared, and the behavior has become smooth in the entire
region.
However, as before, the 4th and 5th Riemann sheets are not connected to each other,
while each is smoothly connected to the 1st Riemann sheet on the real axis.
In fact, the strength function belonging to the 4th Riemann sheet has a pole
(not listed in Table \ref{table1}) in the region of Re $E\geq 21.31$ MeV with Im $E\leq 0$
which is ``hidden'' in the 5th Riemann sheet, but it has almost no effect on the behavior
of the strength function on the real axis. 

Fig.\ref{strpoleE1} shows the total strength function $S_F$ (shown as gray-white shadowgraph) on the real axis
with Im $E=0$ MeV, $S_F^{reg}$ (corresponding to the strength function on the real axis shown in the lower panel of
Fig.\ref{srtGDR3D}) and the contributions of the poles $s_F^{(4)}$, $s_F^{(6)}$, and $s_F^{(7)}$. 
It is numerically confirmed that $S_F^{reg}$, $s_F^{(4)}$, $s_F^{(6)}$, and $s_F^{(7)}$ are the main contributions of $S_F$, {\it i.e.},
the relationship $S_F\approx S_F^{reg}+s_F^{(4)}+s_F^{(6)}+s_F^{(7)}$. 
Looking at the pole contributions individually, $s_F^{(6)}$ is larger than the $S_F$ value around $E=20.72$ MeV,
and contrarily $s_F^{(7)}$ is smaller than the $S_F$ value around $E=22.24$ MeV.
Only $s_F^{(4)}$ has almost the same value of $S_F$ at $E=19.29$ MeV.
Table \ref{table1} shows that the value of $\mbox{Im }r_F^{(g)}/\mbox{Re }r_F^{(g)}$ for pole (7) is a large negative number.
Therefore, $s_F^{(7)}$ has a large negative value in the lower energy region than the peak energy.
So, the value of $S_F$ around $E=20.72$ MeV is smaller than the strength given by $s_F^{(6)}$.
However, around E=19.29 MeV, the contributions from $S_F^{reg}$, $s_F^{(6)}$ and $s_F^{(7)}$ almost cancel each other,
so that the $S_F$ is almost exclusively dominated by the contribution from $s_F^{(4)}$.

These properties are expressed also in the density fluctuations. 
In Fig.\ref{trdE1}, we show the imaginary part of the density fluctuations $\delta\rho_{F,q}(r;E)$ at $E=19.29$, $20.72$, and $22.24$ MeV and
the analysis of the contribution of the poles in the density fluctuations.
The density fluctuations at these three energies all exhibit the typical shape properties of an $E$1 giant resonance, 
but the details are all different.
At $E=19.29$ MeV, the pole (4) contribution $\delta\rho_{F,q}^{(4)}$ dominates the total density fluctuations $\delta\rho_{F,q}$.
Although the density fluctuation $\delta\rho_{F,q}$ at $E=20.72$ MeV are dominated by the contribution of $\delta\rho_{F,q}^{(6)}$,
the contribution of pole $\delta\rho_{F,q}^{(7)}$ is antiphase to the amplitude of $\delta\rho_{F,q}^{(6)}$ for the neutron and
proton components, respectively. This has the effect to reduce the value of the strength function at $E=20.72$ MeV.
The contribution of $\delta\rho_{F,q}^{(4)}$ and $\delta\rho_{F,q}^{reg}$ are also not negligible.
The density fluctuations at $E = 22.24$ MeV are formed by the pole $\delta\rho_{F,q}^{(7)}$ and $\delta\rho_{F,q}^{reg}$
contributions. 

According to Ref.\cite{JostRPA}, the pole (4) originates from the unperturbed pole $\nu[d_{3/2}\otimes(p_{1/2})^{-1}]$
and largely shifts to higher energy regions while obtaining a large width (imaginary part of $E_g$) due to the
effect of residual interactions. 
It is well known from the understanding of RPA calculations using the schematic model with separable interactions
that one of the characteristics of collective modes is the large energy shift from the unperturbed resonance 
peak due to the superposition of many p-h excitations caused by the effect of residual interactions~\cite{RingSchuck}.
Another feature of the $E$1 dipole giant resonance is that the density fluctuations have no nodes and the neutrons
and protons have amplitudes in opposite phases to each other.
Therefore, the pole (4) seems to be the one representing the $E$1 giant dipole resonance since it satisfies all the characteristics of
a collective excitation mode.

However, the aspect of poles (6) and (7) is quite different from that of (4). 
The pole (6) originates from unperturbed pole $\pi[d_{3/2}\otimes(p_{3/2})^{-1}]$, and the position of the pole on the complex energy
plane does not change so much due to residual interactions. However, since the unperturbed strength of $\pi[d_{3/2}\otimes(p_{3/2})^{-1}]$ is not as
large as $s_F^{(6)}$,  the pole (6) is considered to have obtained a large strength due to the effect of the residual interaction. 
The pole (7) originates from the unperturbed pole $\nu[d_{3/2}\otimes(p_{3/2})^{-1}]$ and becomes broadened by the residual
interaction, but with small energy shift.
Since the energy shifts are small in both cases, it can be assumed that the effects on poles (6) and (7) are mainly caused by coupling
with the continuum state.
The difference in effect is almost obviously due to the Coulomb force, because poles (6) and (7) originate from the same unperturbed pole
if there is no Coulomb force.
From the above, although there is no significant difference between poles (6) and (7) and (4) in the total density fluctuation $\delta\rho_{F,q}$,
the detailed properties, structure, and origin of poles (6) and (7) are very different from those of (4), therefore it may be necessary
to distinguish (6) and (7) as excitation modes with different properties from those of (4). 
\begin{figure}[htbp]
  \includegraphics[width=\linewidth]{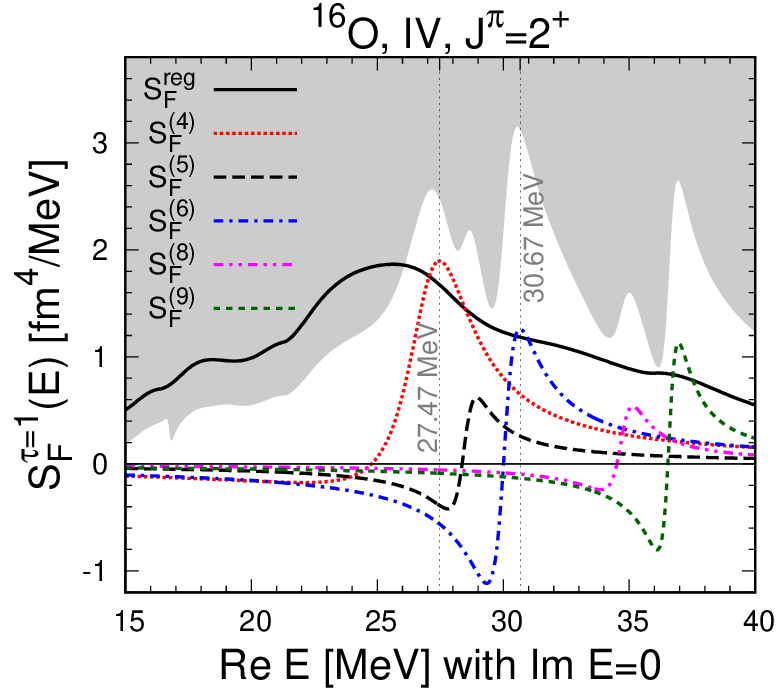}
  \includegraphics[width=\linewidth]{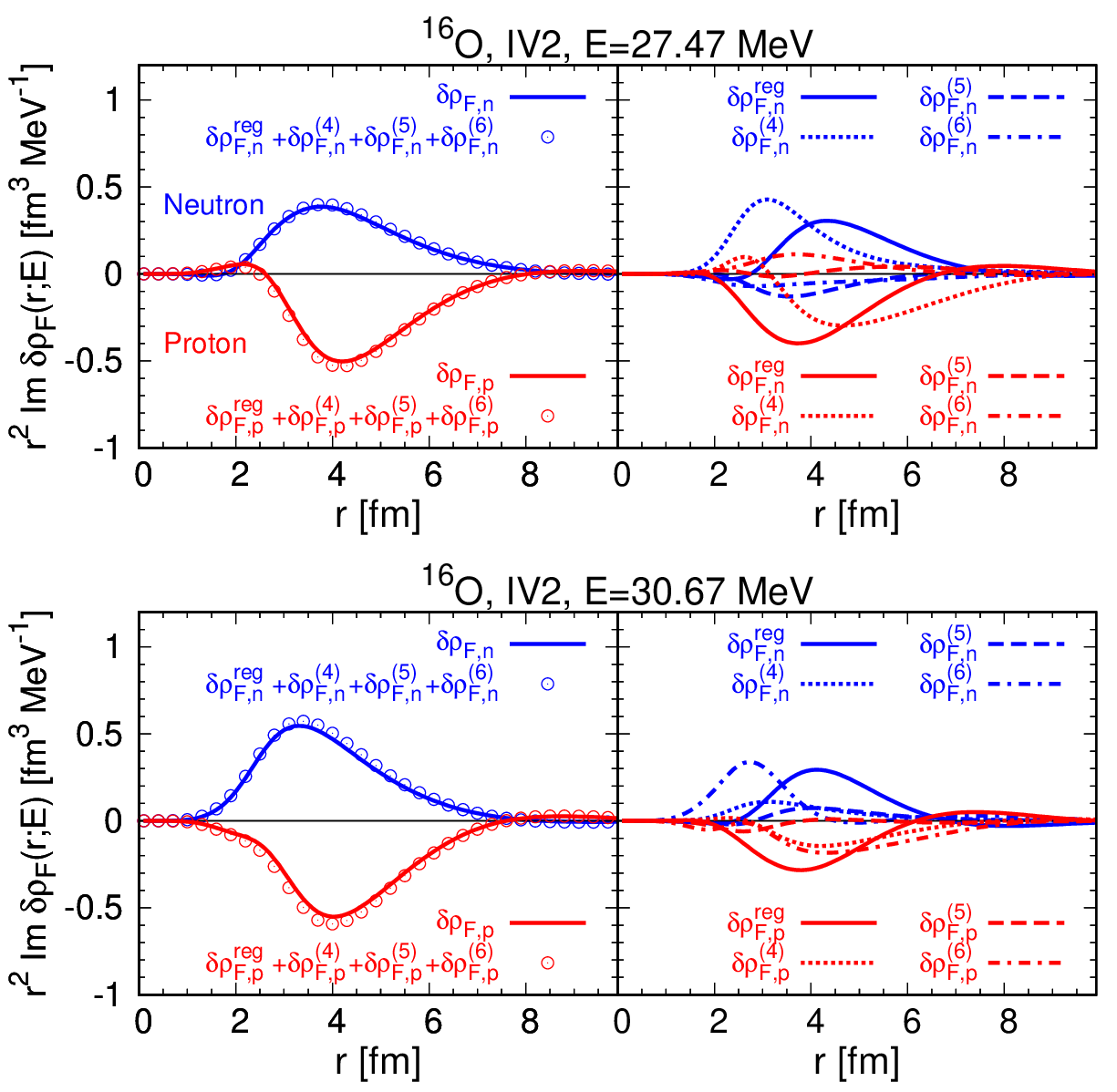}
  \caption{(Color online)
    The Isovector quadrupole strength function for $^{16}$O (upper panel),
    and
    density fluctuations at $E=27.47$(middle panels) and $E=30.67$(lower panels) MeV. Major contribution analysis
    (left panels of middle and lower panels) and decomposition of pole
    coributions (right panels of middle and lower panels).
  }
\label{strtrdIV2}
\end{figure}

\subsection{Isoscalar and Isovector quadrupole excitations}
In Table \ref{table2}, we show the pole $E_g$ of the $^{16}$O quadrupole excitations found on each Riemann sheet,
and their corresponding Re $r_F^{(g)}$, Im $r_F^{(g)}$/Re $r_F^{(g)}$, and $E_{peak}$ in isoscalar and isovector modes,
respectively. 

What stands out in Table \ref{table2} is that poles (1) and (4) have very large Re $r_F^{(g)}$ values compared to the other
poles in isoscalar and isovector modes, respectively.
Since pole (1) has a very small value of Im $r_F^{(g)}$/Re $r_F^{(g)}$, one would expect the contribution of pole (1) to the strength
function to be of Breit-Wigner type shape, and indeed, as shown in the upper panel of Fig.\ref{strpoleL2-1}, the contribution of
pole (1) to the strength function $s_F^{(1)}$ is of Breit-Winer type shape. 
Since pole (1) is isolated from the other poles, it is less affected by $S_F^{reg}$, and $S_F\approx s_F^{(1)}$.
The density fluctuation $\delta\rho_{F,q}$ is also little affected by $\delta\rho_{F,q}^{reg}$, and the density fluctuation $\delta\rho_{F,q}$
at $E=16.76$ MeV is dominated by $\delta\rho_{F,q}^{(1)}$.
And the behavior of the density fluctuations represents the typical isoscalar collective mode properties.
According to Ref.\cite{JostRPA2}, the pole (1) originates from an unperturbed pole of $\pi[f_{7/2}\otimes(p_{3/2})^{-1}]$ and is obtained by a large shift to
lower energies due to residual interactions.
The pole (4) originates from the unperturbed pole $\nu[f_{7/2}\otimes(p_{3/2})^{-1}]$. $\nu[f_{7/2}\otimes(p_{3/2})^{-1}]$ and $\pi[f_{7/2}\otimes(p_{3/2})^{-1}]$
are the same pole if there is no Coulomb force.
In the analysis in Ref.\cite{JostRPA2}, it is confirmed that in the absence of Coulomb force, the unperturbed pole $[f_{7/2}\otimes(p_{3/2})^{-1}]$ splits
into two poles due to the residual interaction, and the isoscalar mode pole shifts to the lower energy side and the isovector mode pole to the higher
energy side. 
Together with these things, pole (4) may also be an isovector collective mode, since pole (4) has a very large value of Re $r_F^{(g)}$.

\begin{figure}[htbp]
  \includegraphics[width=\linewidth]{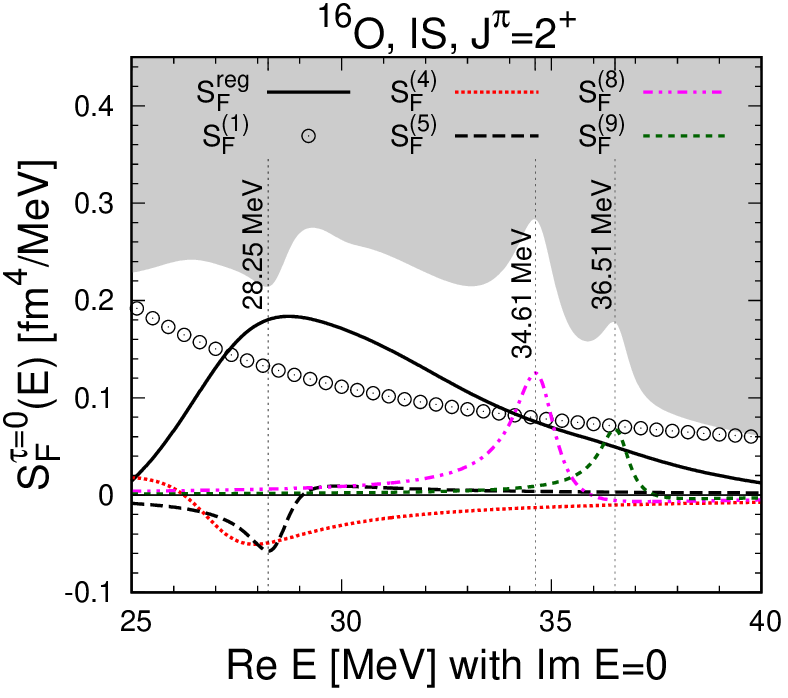}
  \includegraphics[width=\linewidth]{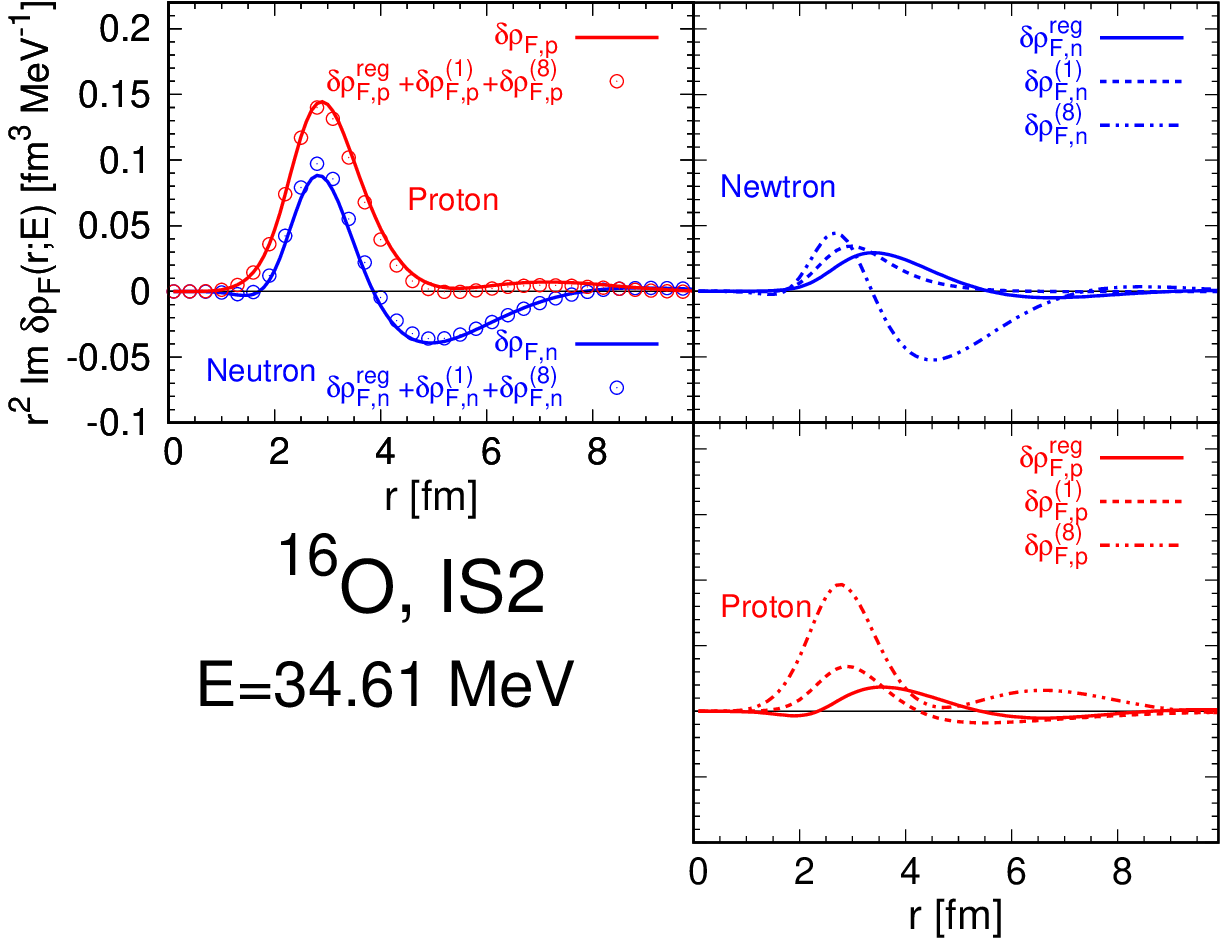}
  \includegraphics[width=\linewidth]{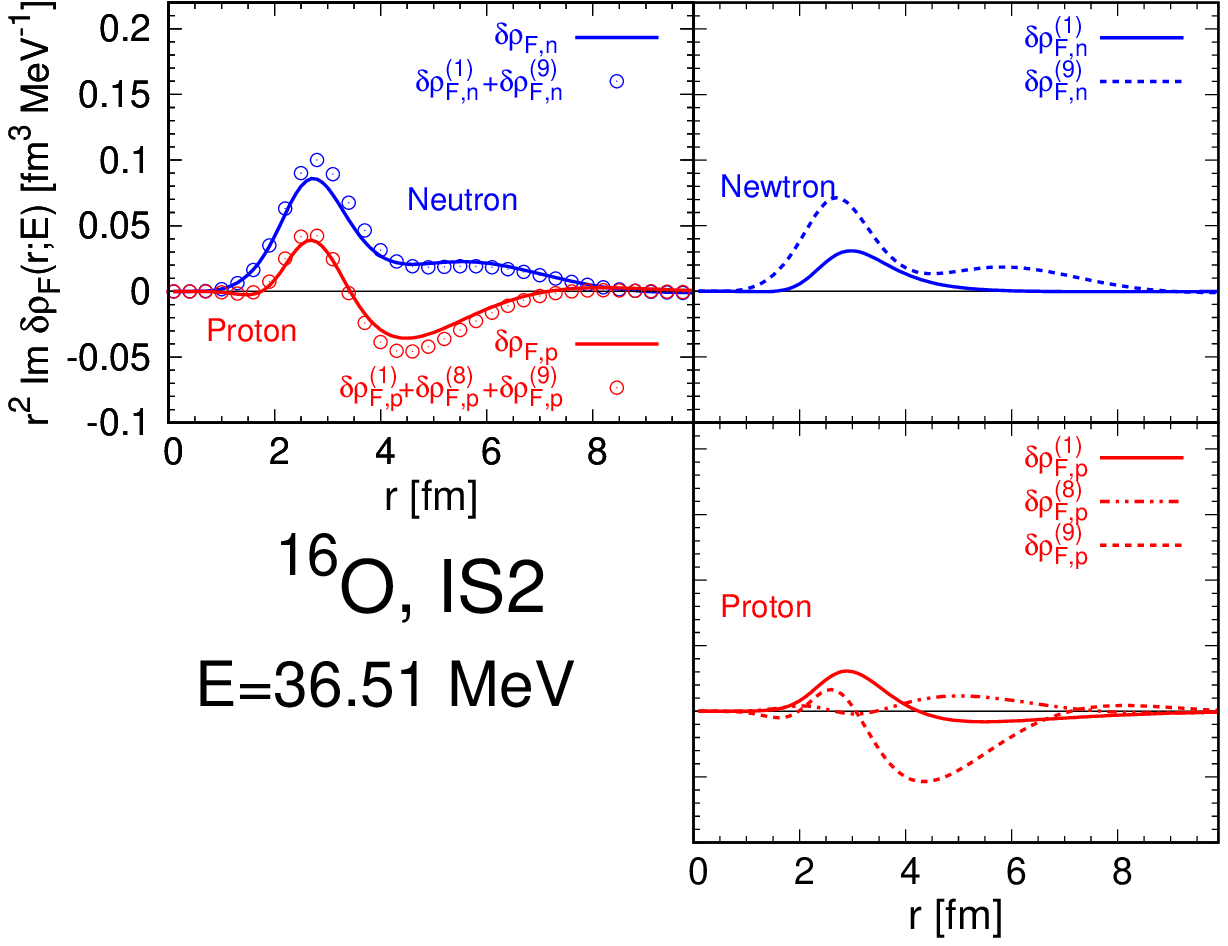}
  \caption{(Color online)
    The same figure with Fig.\ref{strtrdIV2} but for the Isoscalar quadrupole excitation. 
    Density fluctuations at $E=34.61$(middle panels) and $E=36.51$(lower panels) MeV are shown.
  }
  \label{strpoleIS2}
\end{figure}
However, in the case of isovector excitation, the situation is more complicated. 
The strength function for isovector excitation is shown in the upper panel of Fig.\ref{strtrdIV2}. The largest value of the strength
function is not found at the energy near the pole (4), but rather near the pole (6).
This is because, as shown in Table \ref{table2}, all poles in isovector mode have negative Im $r_F^{(g)}$/Re $r_F^{(g)}$ values.
Namely, in the strength function of the isovector mode, the contribution $s_F^{(g)}$ of all poles has an asymmetric shape,
with negative contributions that reduce the strength function at the lower energy side of the poles and positive contributions
that increase the strength function at the higher energy side due to the property of the function $f(X;a)$ Eq.(\ref{Fdef}) in
$s_F^{(g)}$ Eq.(\ref{SFexpand2}) as shown in Fig.\ref{func}.
The value of the strength function near pole (4) is reduced due to the presence of poles (5) and (6) with large negative
Im $r_F^{(g)}$/Re $r_F^{(g)}$ values in the isovector mode in the high energy region near pole (4).
Conversely, the strength function near pole (6) is enhanced by the presence of poles (4) and (5). Therefore, the values of
the peaks corresponding to each pole in the total strength function are reversed at (4) and (6), and are not reflected 
by the value of Re $r_F^{(g)}$. 
Also, as shown by the solid black curve in the upper panel of Fig.\ref{strtrdIV2}, the $S_F^{reg}$ contribution is very
large over the entire energy region.

The middle and lower panels in Fig.\ref{strtrdIV2} show density fluctuations at $E=27.47$ ($E_{peak}$ of pole (4))
and $30.67$ ($E_{peak}$ of pole (6)) MeV for the isovecotor mode.
Looking at the total density fluctuations shown in the left panels, the behavior of the density fluctuations at
both energies seems to be collective mode-like at first glance.
However, the density fluctuations at E=27.47 MeV are formed by the superposition of $\delta\rho_{F,q}^{reg}$ and poles (4), (5), and (6).
Specifically, $\delta\rho_{F,q}^{reg}$ and pole (4) give the main contribution, while poles (5) and (6) contribute in opposite phases and
reduce the value of the strength function, which is consistent with the results found in the strength function analysis.
The density fluctuation at E = 30.67 MeV is also formed by the superposition of $\delta\rho_{F,q}^{reg}$ and poles (4), (5), and (6).
This is also consistent with the results found in the analysis of the strength function. 
These properties are also the same for poles (8) and (9) (Note that pole (7) makes no visible contribution to the strength function), and 
we can see that there is no pole which alone gives the collective mode properties to the behavior of the density
fluctuation for the isovector mode of the quadrupole excitation.
According to Ref.\cite{JostRPA2}, these poles, unlike pole (1), do not shift their energy much and gain width of
resonance due to the effect of residual interaction. 

The nature of these poles in isovector mode is very similar to the poles (7) in $E$1 dipole excitation.
The effect of the residual interaction on the poles is mainly to increase the resonance width rather than the energy shift,
and the contribution of the poles, together with the background contribution ($S_F^{reg}$, $\delta\rho_{F,q}^{reg}$), forms a collective mode-like
structure to the total strength function and density fluctuations.
We think that these properties should be interpreted as indicating that the poles are coupled to the continuum due to the effect of residual
interactions, and that the properties of decay to the continuum state are stronger than the properties of collective motion. 

\begin{figure}[htbp]
\includegraphics[width=\linewidth]{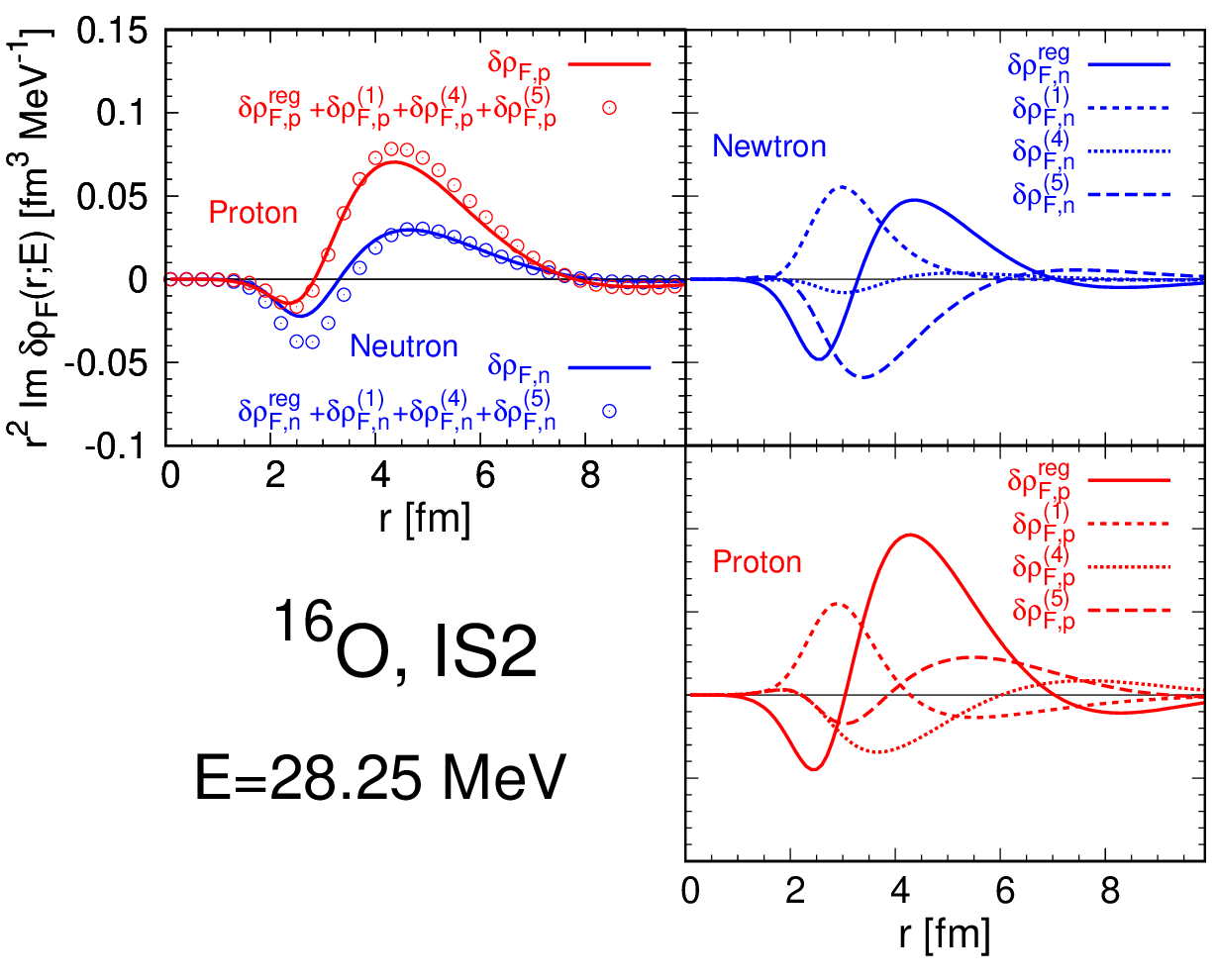}
\caption{(Color online)
  Density fluctuations at $E=28.25$ MeV of isoscalar quadrupole excitations
  in $^{16}$O. Major component analysis (left panel) and decomposition of pole
  components (right panel).
}
\label{trdIS2}
\end{figure}

As seen in the upper panel of Fig.\ref{strpoleIS2}, peaks corresponding to poles (8) and (9)
appear in the strength function around $E=35$ MeV in the isoscalar mode.
Table \ref{table2} shows that the values of $\mbox{Im }r_F^{(g)}/\mbox{Re }r_F^{(g)}$ for the
isoscalar mode of these poles are positive and relatively small.
Since there are no poles with large $\mbox{Im }r_F^{(g)}/\mbox{Re }r_F^{(g)}$ values nearby,
poles (8) and (9) have little influence on each other, and the density fluctuation analysis
shows that each of these peaks clearly shows the character of the corresponding pole. 
As shown in the middle and lower panels of Fig.\ref{strpoleIS2}, the basic shape of the density
fluctuations of the peaks corresponding to poles (8) and (9) are determined by the contributions
from poles (8) and (9), respectively.
At $E=34.61$ MeV, proton density fluctuations have no node and show collective mode-like behavior,
while neutron density fluctuations have a node. The density fluctuation at this energy shows
isoscalar mode-like behavior around $r=2$ to $4$ fm, but it does not show the typical collective
mode characteristics as the mode caused by the pole (1) at $E=16.76$ MeV. 
The peak at $E=36.51$ MeV shows an interchangeable aspect of neutrons and protons in comparison
with the density fluctuation of the peak at $E=34.61$ MeV. Neutron density fluctuations have no
nodes and exhibit collective mode behavior, whereas proton density fluctuations have a node. 
According to Ref.\cite{JostRPA2}, poles (8) and (9) both originate from unperturbed resonance
$[d_{3/2}\otimes(s_{1/2})^{-1}]$ in the absence of Coulomb forces. And they gain width
(imaginary part of pole) without much energy shift due to the effect of residual interactions. 
Perhaps the properties of the poles other than the collective modes presented so far in this
paper may be related to Fano resonance mechanism (Fano effect).
This is because Fano resonance is known as a phenomenon that exhibits asymmetric resonance
spectra caused by the interference between background scattering and resonant scattering processes.
Of course, it is not possible to conclude that Fano resonance exists simply by showing an asymmetric
resonance spectrum. 
Therefore, we next present an example which is difficult to explain without the existence of the
Fano effect. 

Looking at the values of Re $r_F^{(g)}$ for poles (4) and (5) of isoscalar mode in Table \ref{table2},
we can notice that they are negative values. 
Looking at the strength function near $E_{peak}$ ($E = 28.25$ MeV) for pole (5) in the upper
panel of Fig.\ref{strpoleIS2}, we can see that the strength function has a dip.
This dip reflects the fact that the value of Re $r_F^{(g)}$ is negative. This means that poles
(4) and (5) give the contribution to reduce the value of the strength function near $E_{peak}$
of the poles.
Fig.\ref{trdIS2} shows that $\delta\rho_{F,q}^{(5)}$ contributes to reduce the amplitude in the neutron
density fluctuations, while $\delta\rho_{F,q}^{(4)}$ contributes to reduce the amplitude in the proton
density fluctuations. 
In fact, a similar dip can be seen near the $E_{peak}$ of the pole (1) in isovector's strength
function. This reflects the negative value of Re $r_F^{(g)}$ in the isovector mode for pole (1),
as shown in Table \ref{table2}.

In the well-known general understanding of resonance poles, it is believed that when a resonance
pole exists on the complex energy plane, the pole forms a corresponding peak in a physical quantity
such as the strength function or cross section, the real part of the pole gives the peak energy,
and the imaginary part gives the width. 
However, according to Refs.\cite{fano1,fano2,fano3,fano4,fano5, jost-fano,Litvinenko},
when there are interactions which act  between bound or resonant states and continuum,
quantum interferenc effects are induced by the interaction. 
The quantum interference differs from a simple superposition and asymmetrizes the peak structure
of physical quantities such as cross section, corresponding to the resonance poles. 
The asymmetrization of physical quantities caused by such quantum interference effects is called
the Fano effect (or Fano resonance).
It is known that the asymmetric shape exhibited by Fano resonances is described by the Fano formula.
The Fano formula includes the Fano parameter $q$. This parameter represents the ratio of the resonant
scattering amplitude to the background scattering amplitude. 
Depending on the value of $q$, the shape of the resonance spectrum varies significantly: for $q=0$,
the background and resonance scattering are completely canceled, creating a dip that reduces the
cross section to zero at the resonance energy. On the other hand, when $q$ approaches infinity,
the Fano resonance approaches a symmetric Breit-Winer type resonance.
In other words, when the “dip” appears near the resonance energy, it is the most prominent
example of the Fano effect (interference between background scattering and resonance scattering),
and since there are few other effects that produce such a phenomenon, the “dip” may be strong
evidence for the existence of the Fano effect.
Of course, we have not verified the existence of Fano resonance in the excited states of nuclei
in this paper according to Fano's theory, and the relationship between the properties of the pole
residues and Fano resonance is not well understood, so we cannot conclude the possibility of the
existence of Fano resonance in the excited states of nuclei.
However, since Fano resonance is known as a universal phenomena that exists in many-body quantum
open systems as long as there is an interaction between resonance or bound states and continuum,
it is quite possible that it exists in the excited states of nuclei described by the RPA. 

\section{Summary and future perspective}
This paper investigates resonance states in finite quantum many-body systems,
focusing on the $E$1 giant resonance and isoscalar/isovector quadrupole resonances
in $^{16}$O, using the Jost-RPA method. Resonances are analyzed as poles of the
S-matrix, identified as zeros of the Jost function on the complex energy plane.
The study employs the Mittag-Leffler theorem to decompose the RPA response function
into contributions from individual poles and a background term, providing a detailed
understanding of how these poles influence the strength function. 
Although the analysis confirms that some peaks of the strength function are formed
by single poles in $E$1 giant resonances and isoscalar collective modes, it is clear
that many other peaks may be caused by the superposition of contributions from multiple
poles. This highlights the complexity of resonance phenomena, where individual
pole contributions do not necessarily lead straight to an observable peak.

This paper also explores the possibility of quantum interference effects,
similar to Fano resonances, affecting the strength function.
If such quantum interference exists, it may represent a process analogous to
autoionization in atoms, in which the excitation of a nucleus leads to spontaneous
nucleon excitation and subsequent emission or other decay processes. A better
understanding of this would provide insight into the complex interaction between
collective and single-particle behavior in excited nuclei. Therefore,
it is important to clarify the relationship between the properties of the residues
of the poles and the Fano effect.

It is also important to clarify the relationship between the different methods
in order to better understand the detailed nature of the resonance. The approach
used in this paper, which is based on the Mittag-Leffler theorem and methods
based on {\it Berggren}'s ideas~\cite{berggren}, such as the complex scaling
method and the Gamow shell model, are especially important to investigate.
As a first step, we plan to apply the complex scaling method to the Jost-HFB
framework. We have already discussed the possibility of existence of Fano resonances
within the framework of the Jost-HFB method.
If we apply the complex scaling method and the method based on the Mittag-Leffler
theorem within the framework of the Jost-HFB method and compare the results,
we can obtain important insights about the relationship between the complex
scaling method and the method based on the Mittag-Leffler theorem.
On that basis, we believe that a discussion of Fano resonance may provide a
deeper understanding of Fano resonance as well as an understanding of the
resonance analysis method.

So far, we have focused on $^{16}$O as the target nucleus in order to establish a
resonance analysis method using the Jost-RPA method. However, in order to obtain
a systematic understanding of resonances based on the Jost-RPA method, we are
planning to apply the method to other heavier nuclei and various excited states
in order to perform a systematic analysis.
For realistic calculations with comparisons with experiments in mind, it may be
necessary to consider performing self-consistent calculations using effective
nuclear forces (such as the Skyrme force). 

Although we cannot be sure without actual calculations, in principle the Jost-RPA
method can be considered a unified method for nuclear structure and nuclear reactions,
including elastic, inelastic, and single nucleon-capture channels, and may be applicable
to nucleon scattering targeting even-odd and odd-even nuclei.
Since the intermediate state of the reaction is required to be a double closed-shell
nucleus for nucleon scattering using the Jost-RPA method, $n+^{207}$Pb would be a good
candidate for the first calculation.
There are a many interesting things to investigate, such as the influence of the excited
resonance state of $^{208}$Pb via in the intermediate state and the optical potentials that
could be derived within the framework of the Jost-RPA method and so on. 

However, since the scale of the
calculation will be very large and the analysis work is expected to be enormous,
we plan to work on the above issues step by step in the future.

\end{document}